\documentclass[12pt]{article}
\usepackage{amsmath}
\usepackage[dvips]{graphicx}
\usepackage{epsfig}
\usepackage{amsmath}
\usepackage{amssymb}

\newcommand{\bm}{\begin{multiline}}
\newcommand{\beq}{\begin{equation}}
\newcommand{\eeq}{\end{equation}}
\newcommand{\beqs}{\begin{eqnarray}}
\newcommand{\eeqs}{\end{eqnarray}}

\begin{document}

\thispagestyle{empty}

\hfill{}

\hfill{}

\hfill{}

\vspace{32pt}

\begin{center}

\textbf{\Large DYNAMICAL FLUID-TYPE UNIVERSE $\vspace{10pt}$ SCENARIO WITH DUST AND RADIATION}

\vspace{48pt}

\textbf{ Denisa-Andreea Mihu}\footnote{E-mail: \texttt{denisa.mihu0@gmail.com}}

\vspace*{0.2cm}

\textit{Faculty of Physics, ``Al. I. Cuza" University}\\[0pt]
\textit{11 Bd. Carol I, Iasi, 700506, Romania}\\[.5em]

\end{center}

\vspace{30pt}

\begin{abstract}

		Within the context of a cosmic space whose energy source is modeled with a perfect fluid, a uniform model of Universe based on a standard FRW cosmology containing decoupled mixed matter sources namely stiff matter and cosmic dust together with a positive cosmological constant, has been studied. Within the scenario of a $k=0-$ spatially-flat geometry, we analysed the geometrodynamics of the considered theoretical cosmology. For the model with an added cosmological constant, the main scope was to point out the effects of it on the model dynamics. In this last case, the thermodynamics of the model was also considered together with the relation between the cosmological energy density and fluid pressure in terms of the inverse function of equation of state.  

\end{abstract}

\vspace{32pt}

\setcounter{footnote}{0}

\newpage

\section{Introduction}

		 Nowadays, an almost ultra-new trend in cosmology is revealing itself through consistent investigations focusing on the mixed matter type universes, most of them elaborated in the frame of Einstein theory of general relativity. Today, it is well-known that matter composition of the Universe is not a single one instead, it is represented by 5\% baryonic matter, 20\% dark matter and 75\% comes in terms of dark energy \cite{Binney:2008sr}. According to the observations, we are provided with the conclusion that within the present epoch the main contribution to the energy density is reserved to the cosmological constant, the most trivial and appealing candidate to dark energy. It proves that this dark energy component is the one which drives the expansion of the universe at an accelerated rate. The role of the cosmological constant in the evolution of the Universe cannot be denied at all as it was shown that inserting the cosmological term in the theory of the Big - Bang standard model of Universe results in the correlation of this theoretical framework with the large scale experimental observations on the distribution of galaxies and clusters, with WMAP measurements on the fluctuations detected in CMB radiation and also with the properties which were discovered on the level of the X-ray emission clusters \cite{Krauss:1995sr}. Thus, the cosmological constant is the most trivial case in accordance with the existing data, more general approaches to dark energy or modified gravity leading to weaker constraints on the other parameters. 

		 In order to shed light into the nature of the Universe, into its features and its properties at megascale, modern cosmology brings into attention a wide range of matter sources, from common ones to exotic phases – particularly dark matter, this a mysterious state of matter, exotic dust, but not only: Bose-Einstein condensates or quark-gluon plasma, these as states of matter that are not commonly encountered, but whose properties are entirely situated in the realm of mainstream physics; with the same kind of properties as the last, to be mentioned exotic baryons representing hypothetical particles; so far, the only observed baryons are pentaquarks $P_{c}^{+}(4380)$ and $P_{c}^{+}(4450)$ that were put into evidence by LHCb mission in 2015 \cite{Aaij:2015sr} $-$ which incorporate the expansion phenomenon of the universe within the solutions of Einstein's field equations. It is well-known that the expansion rate is intimately related to the types of matter and energy that the Universe contains and, particularly, the position of the total energy density related to the so-called critical density. As the universe undergoes a continuous evolutive state with progressive phases in which new forms of matter are generated, with various stages that affect the stages that preceded them and consequently the behavior of the universe as a whole, a theoretical analysis of the cosmologies that consider a mixed composition of matter seem legitimate and, in the same time, very appealing. More than this, one can obtain a more accurate and closer description of the Universe in its present state or there can be configured the transitions between one epoch to another or, more recently, it might be revealed how these matter sources intermingle to generate new physically interesting dynamics. All these ideas and, mostly, this last particular case is an appealing subject these days which ignited our interests and it is staying beyond the motivation of the present paper. 

		Furthermore, the isotropy and homogeneity which characterizes the large scale universe are embedded in the fluid - filled FLRW evolving cosmological models \cite{Hawking:1973sr} which are locally isotropic everywhere and consequently also spatially homogeneous. FLRW cosmologies are viewed as the standard models at the basis of modern cosmology \cite{Weinberg:1972sr}. The detection of the CMB radiation generated within a very early and hectic phase as a label of the relic radiation come to support this idea. According to the data collected by the COBE mission, it appears that the CMB radiation is very smooth to at least one part in $10^{5}$\cite{Bennett:1996sr}, possessing a black body spectrum whose temperature isotropy constitutes a direct probe in favor of the physical constraint of homogeneity and isotropy of the Universe at its megagalactic scale. Observations on temperature variations in CMB indicate that after leaving the radiation era, the geometry of the Universe is described in terms of a spatially-flat FRW \cite{Bernardis:2000sr}. We specify that the large-scale homogeneity as a characteristic of the observable universe founds a powerful explanation within the inflation scenario \cite{Guth:1981sr}. Thus, these aspects come to support FRW-type cosmologies in the context of the fluidic formalism.

		Considering the exposed aspects, we dedicate this paper to theoretical investigations performed on a combined type of matter universe in a $k=0-$ FRW spatially flat geometry. A similar approach was applied by us in a previous study \cite{Dariescu:2016sr} were we developed and analysed a mixed matter cosmology with non-interacting ideal fluids namely, stiff matter, dust and a positive cosmological constant. In the first stage of our present study, we consider the universe content modeled by a perfect fluid comprising dust and radiation and we attempt to find out the solution corresponding to this model of Universe. In the second stage, we add one more component, namely a non-vanishing cosmological constant in order to study the effects of this component upon the dynamics of the considered model. We specify that the species of fluid involved are treated as non-interacting matter sources. Historically, models with decoupled matter and radiation start with the one elaborated by Lemaitre \cite{Lemaitre:1931sr} followed by Stabell \cite{Stabell:1968sr} and McIntosh \cite{Nume:1}. In spite of this assumption we used to work with, it seems that cosmologies which involve connected matter and radiation might be more accurate and more closer to physical reality. For instance, we indicate such treatments in the studies of McIntosh \cite{Cbg: 1968sr}, May and McVittie \cite{May:1970sr}. While McVittie investigations focused only on the flat models $(k=0)$ with a non-linear time-dependent equation of state parameter, May and McVittie considered models of universe in the context of the three types of geometries $(k=0, \pm1)$,  operating with a rate of change in the scale factor described through the relation $da/ dt=cg(\varphi)$ with $\varphi$ a function of time and $c$  denoting the speed of light.
		
		In this study, we analyse the geometrodynamics together with the main features of a theoretical model of universe with a combined matter composition formed with decoupled dust and radiation treated as perfect fluids. At the basis of our choice for the togetherness of the radiation and dust species lies the phenomenology of the influence of radiation pressure on the cosmic dust mostly at the level of galaxies. For a long period way back in the past, this issue is a very acute and debatable one with some of the involved problematics still unelucidated at the time we discuss them. In view of these, in \cite{Morgan:1942sr}, one will find a detailed presentation of the problems which has its root in the fact that cosmic dust comes with a consistent contribution to the mass of the galaxies. On the other hand, we noticed that cosmology with decoupled dust and radiation sources made the subject of some other theoretical investigative models \cite{PNeto:2005sr, CHAVANIS:2015sr}. 

		Within the history of cosmological solutions of Einstein field equations, dust has been occupying a relevant place. This type of matter has been present starting with very early phases of the universe, some galaxies being identified as containing dust particles which could lead to the formation of planets \cite{Watson:2015sr}. Besides this, as reported by many studies, it appears that within our Universe the dust medium is regnant as it is intensively generated in various chemical and physical processes possessing its specific function in interstellar surface chemistry \cite{Williams:2002sr}. In \cite{Brandt:2014sr}, it was suggested that dusty remnants produce some kind of phenomena termed as meteor showers. According to this study, the solar radiative impact on the surface layers of comets activates the process of sublimation of the ices which results in the generation of gas and liberate dust particles. The latter arrange in a specific structure called dust tail. 

		With reference to the radiative component, this component has played an important role in those early phases of cosmogenesis, at that time, the largest quantity of energy in the Universe being in the form of radiation, i.e. in the form of the subatomic constituients which undergo a  relativistic motion, mostly it is about photons and neutrinos. The radiative dominant epoch is encountered at about 300000 years after the Big-Bang event, after this time the universe proving to become itself transparent to photons. 

		In the same time, we performed investigations on the same model in the presence of a positive cosmological constant. In the beginnings of cosmological studies, consideration was given to models with no cosmological term ($\Lambda$=0). Also, there can be identified studies on Friedmann models with pressureless content and a non-vanishing cosmological constant \cite{H. P. Robertson}. In the early stages of these cosmological researches, less considerations have been attributed to more general Lemaître models with a given relation connecting the pressure and the energy density in terms of an equation of state $p=p(\rho)$ and  ($\Lambda\ne0$) \cite{G. C. McVittie:1965sr}. Afterwards, in the view of the age problem \cite{Krauss:1995sr, Turner:1997sr}, there were questioned the effects of a non-zero cosmological constant. With reference to the magnitude of this cosmological constant, here we will work in the frame of a small one.  Recently outstanding experimental measurements of two independent groups, High-Z Supernova Team and the Supernova Cosmological Project \cite{Schmidt:1998sr} strongly favor a small positive value for the cosmological constant.
As for large enough values of it, it proves that the age of the universe can achieve an infinite value, this due to the fact that backward in time $a(t)$ possesses a non-vanishing inferior threshold $a_{min}$. This type of 'big bangless' oscillating models prove to be acknowledged for over sixty years \cite{H. P. Robertson:1933sr}. Nonetheless, on purposes that the required values of $\Lambda$ do not match the observational data \cite{G. F. R. Ellis:1984sr}, they have been ignored being labeled as unphysical.
		
		More recently, it was proposed an interesting theoretical framework for modeling dark matter namely `mimetic dark matter' \cite{Chamseddin:2013sr}. In addition, various models have been designed that describe a dark energy fluidic universe \cite{Nojiri:2004pf}.
 The dark energy sector possesses a negative pressure being responsible for the today acceleration of the Hubble expansion. In terms of possible candidates are proposed the cosmological constant, the vacuum density of quantum field theory or a dynamical component characterized by time dependent energy density and spatial distribution due to different kinds of fields such as a canonical scalar field termed quintessence ($w>-1$) \cite{Ratra:1988sr} or a phantom field ($w<-1$) which is nothing else but a scalar field described with a kinetic term which comes with a negative sign \cite{Caldwell:2002sr} or a quintom \cite{Guo:2005sr}. There are also other alternatives such as ghost condensate \cite{Piazza:2004sr} or k-essence \cite{Chiba:2000sr} or quite recent attempts to model its effects on the acceleration of the Universe with an exotic type fluid with negative pressure known as Chaplygin gas \cite{Bento:2002sr}, but all these alternatives are going beyond the aim of our paper. A detailed review on various dark energy cosmologies and correlations with the data analysis or cosmography tests are to be found in \cite{Bamba:2012cp}.
		
		Our approach is inspired by the work of Chavanis  \cite{CHAVANIS:2015sr} which we shall extend in order to gain more insights into the model’ dynamics. Let us specify that a theoretically different approach related to the one we will tackle in the present paper is found in the study \cite{PNeto:2005sr} where, within the frame of quantum minisuperspace theory, the authors analysed a series the features of FLRW universes with single and two non-interacting fluids: dust and radiation. The paper theoretically proves that within a combined dust-radiation universe, the matter in the form of dust can be created as a quantum effect such that an exotic type of dust matter arises and the universe can suffer a transition from an exotic dust matter phase to a conventional dust matter era. The study emphasizes that when quantum effects come into play, the radiation also possesses an exotic character, leading to the formation of bounces.
It proves that the fluidic scenario, approach that we will also consider  for here present study, might have physical applications with a quite interesting side: exotic fluids causing cosmological bounces and avoiding cosmological singularities \cite{Peter:2001sr}, these also being relevant, for example, for the creation of worm-holes \cite{Morris:1988sr}.

\section{Model Dynamics within the context of fluidic energy source}

	       The line element in comoving spherical coordinates for the quadridimensional homogeneous and isotropic flat space with constant local curvature is given by the $k=0-$ Friedmann-Robertson-Walker (FRW) metric	
 \begin{eqnarray}  \label{newinitialmetric1}
ds^{2} = a(t)^2[dr^2+r^2d\Omega^2]-dt^2,
\end{eqnarray}
where $d\Omega^2=d\theta^2+\sin^2 \theta d\varphi^2$ defines the metric of the unit sphere $S^2$. The physical quantity $a(t):\Re \rightarrow \Re_{+} $, $a(t)=a_{0}e^{f(t)}$ defines the scale factor, a function of the global time coordinate namely the cosmic time denoted here by $t$. 

Within the framework of a uniform perfect fluid of isotropic pressure with the energy-momentum tensor components described by $T_{\alpha\alpha}=p,  T_{44}=\rho$, the Einstein’s system of equations, $G_{ab}+\eta_{ab}\Lambda=\kappa T_{ab}$ , admits the explicit representation        

\begin{eqnarray}
2 {\ddot a\over a}+{ \dot a^2 \over a^2}-\Lambda&=&-\kappa p \nonumber\\ \label{einstein1}
3{ \dot a^2 \over a^2}-\Lambda&=&\kappa \rho,  
\label{einstein2}
\end{eqnarray}
where  $\kappa={8\pi G}/ {c^4}$ and $\Lambda$ is the positive cosmological constant.	

The second relation in (\ref{einstein2}), 

\begin{equation} 
\label{Friedmann}
H^2={\kappa \over 3}\rho + {\Lambda \over 3},
\end{equation}
where $H$ is the Hubble function defined as $H={\dot a / a}$ , is known as the Friedmann equation for $k=0-$ FRW Universe and it will be of principal use in the present study as it correlates two of the most essential cosmological parameters responsible for the dynamics of the Universe: the total energy density $\rho$ of the perfect fluid filled - Universe and the Hubble expansion parameter $H$.

	As in the standard $\Lambda CDM$ model which treats the four matter components, namely radiation, baryonic matter, dark matter and dark energy, as independent constituients with their proper equation of state ($p={\rho/3}$ for radiation), ($p=0$ for baryonic matter and dark matter), ($p=-\rho c^2$ for dark energy), the energy density of the model being described through the summation of the energy densities characterizing each species separately, our cosmological model containing a combined matter source made of non-miscible radiation and dust matter will have the total energy density given by

\begin{equation}
\label{totalED}
\rho=\rho_{01}{\left({a_0 \over a}\right)^3} +\rho_{02} {\left({a_0\over a}\right)^4}={3 \over \kappa}\left({\beta \over a^3}+{\gamma \over a^4}\right). 
\end{equation}
We specify that there were performed the following notations $\beta={\kappa \over 3}\rho_{01}{a_0}^3$ $($cosmic   dust$ )$, $\gamma={\kappa \over 3}\rho_{02}{a_0}^4$ $($radiation$)$, where the zero index corresponds to the present day values. 

With this result, the general Friedmann equation (\ref{Friedmann}) becomes equivalent with the differential equation

\begin{equation}\label{adot}
{da \over dt}= \sqrt{{\beta \over a}+{\gamma \over a^2}}.
\end{equation}
Performing the integration of the first order differential equation (\ref{adot})  with imposing the origin condition $a_{i}(t=t_{i}=0)=0$, we arrive at the equation

\begin{equation} \label{cubic}
a^3-3{\gamma \over \beta} a^2={9 \over 4} \beta t^2-6{\gamma ^{3/2} \over \beta}t
\end{equation}
which is a cubic equation in the scale factor’ variable. To be mentioned that our result can be put into correspondence with the one obtained in \cite{CHAVANIS:2015sr} for the same choice of Universe composition. 

	Before proceeding to an analysis of the solutions of this equation, we will focus on deducing the expression of the other essential cosmological quantities.
First of all, it is worth seeing that at the primordial cosmic singularity, $a(t=0)=0$, the energy density  (\ref{totalED})  is blowing up as it goes to infinity $(\rho \to +\infty)$.
From (\ref{adot}), one is able to compute the Hubble parameter as dependent on the scale function:

\begin{equation}
H(a)=\sqrt{{\beta \over a^3}+{\gamma \over a^4}}.
\end{equation}
Further on, inserting these results into the Einstein equation one can deduce the evolution of the pressure with respect to the scale factor as described by

\begin{equation}\label{Press_a}
p(a)={\gamma \over \kappa a^4},
\end{equation}
formula which together with (\ref{totalED}) allows the finding of a non-linear dependence of the cosmological energy density on the fluid pressure:

\begin{equation} \label{inversefct}
\rho(p)=3\left[{\beta(\kappa{\gamma^3})^ {-1/4}{p}^{3/4}+p}\right].
\end{equation}
This result is the inverse function of the EoS of our present model, $p(\rho)$, the latter possessing a very intricate mathematical representation.

To be noticed that the dependence (\ref{Press_a}) of the fluid pressure with respect to the scale function emphasizes a radiative behaviour.
	
From the model’ time dependent EoS, $p=w(t)\rho$, we compute the effective EoS parameter 

\begin{equation}
w(a)={\gamma \over {3(\beta a+\gamma)}} \nonumber\\
\end{equation}
which results to decrease with the ratio $\beta \over \gamma $ of the dust species over the radiation component. 
Also, as the scale function approaches the primordial cosmological singularity, $a\to 0$, the EoS parameter goes to a constant, $w \to {1 \over 3}$. This means that the early universe has been dominated by the radiative component. Obviously, the same $w$ asymptotic tendency appears if ${\beta \over \gamma} \to 0 $ (a dominance of the radiative component over the dust species). In the last years, time-variation of the EoS parameter has been under ardent investigations and discussions, for instance, within the context of the models with viscous fluids \cite{Rahaman:2006sr} or quintessence models with scalar fields \cite{Caldwell:1998sr}, and many techniques based on expressional data for recovering the physical quantity $w(t)$ have been originated \cite{Sahni Starob:2006sr}. In addition to this, experimental measurements were reviewed in order to explore the evolution of this parameter with respect to the cosmic time \cite{Sahni:2008sr} and the references therein. For further thermodynamics insights on cosmologies with time dependent equation of state parameter we suggest \cite{Magueijo:2008sr}. 

		Now, knowing that our model reflects in its early phases of the cosmogenesis a radiation era which manifests its dominance for times compatible with ${a(t) \over a(t_0)}<10^{-4}$, our result appears to be justified mostly that we worked out within the primordial conditions $a(t=0)=0$. Considering this idea, we can state that regardless of the cosmology that suits best for the actual universe, we can be admit with confidence that we know the time evolution of the scale factor within the first few tens of thousands of years after the primordial cosmic explosion (Big-Bang): $a(t) \sim t^{1/2}$.

		Last but not least, we compute the acceleration parameter by using its well-known definition,

\begin{equation}
q(t)=-{{\ddot a}a \over {a^2}} \nonumber\\
\end{equation}
leading to

\begin{equation}  \label{accparam_a}
q(a)={{2\gamma +\beta a}\over 2(\beta a+\gamma)}>0.
\end{equation}

In view of this result, taking into account that $a(t)$ is itself a positive defined quantity, we conclude that our cosmological model portrays a decelerating Universe (i.e., $q(t)>0$).

At this point, we will return to the equation (\ref{cubic}) and by calling down the theory of third degree equation \cite{Irving:2004sr}, this provides us a framework in which one can discuss the nature of roots of the cubic equation (\ref{cubic}) as a function of model’ parameters. In this respect, in what follows, we will dedicate ourselves to this detailed analysis.
			
Hence, the discriminant for the equation (\ref{cubic}) is 

\begin{equation} \label{DELTA}
\Delta=-27Q^2-108{\gamma^3 \over \beta^3}Q,
\end{equation}
where one should take into account the notation

\begin{equation} \label{notQ}
Q \equiv {9 \over 4}\beta t^2-6 {\gamma^{3/2} \over \beta} t={9 \over 4}\beta t(t-t_{*})
\end{equation} 
with $ t_{*}={8 \over 3}{\gamma^{3/2} \over \beta^2}$.
Depending on the signs of both the discriminant (\ref{DELTA}) and of the expression in (\ref{notQ}), different situations distinguish. As a next step in our investigations, we shall dedicate to these situations our mathematical and physical treatment.  

	\vspace{15pt}
	i. $\Delta<0 $ and $Q>0$

	\vspace{15pt}

In this case, equation (\ref{cubic}) possesses one real root and two complex conjugate roots. Because the scale factor has to be real, we shall select the corresponding solution. This solution is given by the algebraic construction

\begin{eqnarray} \label{SOLi}
a(t) \equiv a(Q)&=&{\gamma \over \beta}+{{2^{1/3}\gamma^2} \over \beta} \left({\beta^3Q+2\gamma^3-\beta^3Q\sqrt{1+{{4\gamma^3} \over {\beta^3Q}}}}\right)^{-1/3}+ \nonumber\\
		    &&{2^{-1/3} \over \beta} \left({\beta^3Q+2\gamma^3-\beta^3Q\sqrt{1
		    +{{4\gamma^3} \over {\beta^3Q}}}}\right)^{1/3}
\end{eqnarray}
with $Q$  representing the time dependent function defined in (\ref{notQ}).

\rm In the very late phases of the Universe’ evolution, phases that are compatible with small values of the ratio ${4\gamma^3} \over {\beta^3Q}$ , i.e. high values of time variable (and, consequently, high values of $Q$ ), one can perform a series expansion of the square root and, in this situation, the scale factor admits the asymptotic representation below:

\begin{equation}
a_{t \to  \infty} \approx {\gamma \over \beta} + Q^{1/3}+{\gamma \over \beta} Q^{-1/3}  \approx Q^{1/3}.
\end{equation}
Recalling (\ref{SOLi}), for large values of the time variable, i.e. $t \gg t_{*}$,  we have the behavior $Q \approx {{9 \over 4} \beta t^2}$ from where we find that the radius of the universe increases algebraically as

\begin{equation}
a \approx \left({{9 \over 4} \beta t^2}\right)^{1/3} \sim t^{2/3}.
\end{equation}
A closer look to this result leads us to the conclusion that, in those very late times of the Universe’ evolution, we no longer have a ‘unified’ description with both phases, cosmic dust and relativistic matter, contributing in their proper way, instead we find a dust-like Universe similar to the one described by the EdS universe. As one expected, in the late times, the radiative component does not manifest any effects.
	
		In what concerns the other essential cosmological parameters, we deduced their evolution in the very late universe. Thus, for the Hubble function we find

\begin{equation}
H_{t \to  \infty} \approx {2 \over 3t} \sqrt{1+{\gamma \over {\beta^{4/3}}}\left({4 \over 9}\right)^{1/3}t^{-2/3}}
\end{equation}
which behaves asymptotically as $H \sim 2/3t$.
	
	For the cosmological energy density, its variation results to be described through the following algebraic representation 

\begin{equation}
\rho_{t \to  \infty} \approx {{1 \over \kappa}\bigg[{4\over {3t^2}}+\left({2\over 3}\right)^{5/3}{{{2\gamma}\over {\beta^{4/3}}}{t^{-8/3}}}}\bigg] \approx {4 \over {3\kappa t^2}} \to 0,
\end{equation}
while for the pressure time evolution we have

\begin{equation}
p \approx {{\gamma} \over{\kappa \beta^{4/3}}}\left({4 \over 9}\right)^{4\over 3}t^{-8/3} \to 0.
\end{equation}
To be seen that, at late times, both the energy density and the pressure are vanishing. 

	\vspace{15pt}
	ii. $\Delta<0 $ and $Q<0$  $(t<t_{*})$

	\vspace{15pt}

In this case, the solution is the same as in the previous one, but it proves that this situation cannot take place as the two inequalities cannot be valid simultaneously. If that were the case, then one will arrive at the inequality ${{4\gamma^3} \over {\beta^2}}-{\mid Q \mid}<0$, which with considering (12) and ${t_{*}}={8 \over 3}{\gamma^{3/2} \over{\beta^{2}}}$, becomes equivalent with ${(2t-t_{*})}^2<0$ which obviously is false. On the other hand, we must emphasize that the condition $Q<0$ implying $t<t_{*}$ illustrates a time limited universe with its limit set by the parameters of the model. This case does not constitute a physical one.

	\vspace{15pt}
	iii. $\Delta=0 $ and $Q<0$  $(t<t_{*})$

	\vspace{15pt}

For this combination, it results that  $Q=-{4 \gamma^3 \over \beta^3}$ which transforms our cubic equation into the following configuration

\begin{equation}  \label{EQiii}
a^3-{3\gamma \over \beta}a^2+4 {\gamma^3 \over \beta^3}=0.
\end{equation}
The new form (\ref{EQiii}) admits a static solution $a_{1}=a_{2}={2 \gamma \over \beta}$ corresponding to an unevolving universe (static universe) and a root in terms of $a_{3}={\gamma^2 \over \beta^2}Q+{3\gamma \over \beta} $. If we ask for $a_{3}$ to be a positive physical quantity, one will have to deal, as in the previous case, with a time limited universe characterized by the parametric interval $t \in \left(0, {1 \over 4}t_{*}\right) \cup \left({3 \over 4}t_{*}, t_{*}\right)$ or  $t \in \left(0, {2 \gamma \sqrt{\gamma}\over 3\beta^2}\right) \cup \left({2\gamma \sqrt{\gamma}\over \beta^2}, {{8\gamma \sqrt{\gamma}}\over {3\beta^2}}\right)$. Obviously, neither this case cannot be considered to be a physical one. 

\section{\rm \bf  Thermodynamics of the Model }

		From the first principle of thermodynamics, we recall the thermodynamical equation \cite{Weinberg:1972sr}

\begin{equation}
{dp \over dT}={{1 \over T} (\rho+p)}
\end{equation}
which coupled with the inverse function of the EoS (\ref{inversefct}) can be integrated in order to obtain the dependence of pressure with respect to temperature. In this way, it was deduced the relation

\begin{equation}  \label{pressTemp}
{p(T)}={\left({3\beta \over 4}\right)^{4} {1 \over {\kappa\gamma^3}}}\left({{T \over {T_0}}-1}\right)^{4},
\end{equation}
where the integration constant has been associated to the minimum temperature $T_{0}$.

Having the expression (\ref{pressTemp}) and making use of it on the inverse function of the EoS (\ref{inversefct}), this yields the thermodynamics which governs the energy density:

\begin{equation}    \label{rhoTemp}
{\rho(T)}={\left({3\beta \over 4}\right)^{4} {1 \over {\kappa\gamma^3}}}\left({{3T \over {T_0}}+1}\right)\left({{T \over {T_0}}-1}\right)^{3}.
\end{equation}
For $T \gg T_{0}$, condition compatible with the early phases of cosmogenesis, the dependence (\ref{rhoTemp}) reduces to a Stefan Boltzmann law type dependence, $\rho(T)\sim{\left({3\beta \over 4}\right)^{4} {3 \over {\kappa\gamma^3T_{0}^{4}}}}T^{4}$, where the pre-factor might be identified with the Planck temperature $T_{Planck}\equiv{\left({3\beta \over 4}\right)^{4} {3 \over {\kappa\gamma^3T_{0}^{4}}}}$. As a remark, we find this result compatible with the radiative epoch that we found above through the equation of state parameter $w$, to characterize the early times of this cosmology.
A similar thermodynamical variation law may be found in \cite{Chava:2012sxp} in the frame of a radiation dominated-cosmology, free of singularity, describing the early universe modeled with a polytropic equation of state of the form $p={1 \over 3}\left({1-{4\rho \over \rho_{P}}}\right)\rho c^2$, with $\rho_{P} \equiv \rho_{Planck}$ representing a constant of integration which proves plausible to be regarded as the superior limit of the density in terms of the well-known Planck density ($\rho_{Planck}=5.16 \cdot 10^{99} g / m^{3}$).
	
	Operating at a trivial level with (\ref{Press_a}) and (\ref{pressTemp}), we obtain the thermodynamic variation characterizing the scale function:

\begin{equation}  \label{aTemp}
{a(T)}={{4\gamma} \over 3\beta}\left({{T \over {T_0}}-1}\right)^{-1}.
\end{equation}
From the relation (\ref{aTemp}) it can be easily inferred that as the Universe evolves, its temperature drops towards its minimum value $T_{0}$ according to the implication $a\approx t^{2/3} \to T=T_{0}$. Also, at these late times, one may note that both the pressure and the energy density tend to vanish ($p \to 0$; $\rho \to {0}$).

Finally, one is able to compute the entropy of the universe 

\begin{equation}
{S(T)}={a^{3} \over T}\left(\rho+p\right)={{3\beta} \over {\kappa T_{0}}},
\end{equation}
resulting a constant value which is in alignment with the theoretical framework’ universe characteristics of homogeneity and isotropy characterizing our model. Also, to be noticed that it is expressed in terms of the parameter characterizing the dust component. We remind that $T_{0}$ represents the lower threshold of the temperature associated to the integration constant.

\section{The Model with a Cosmological Constant}

		In the section we shall dedicate ourselves to the study of the effects that a cosmological constant can generate if inserted in the model we already discussed. Thus, if one takes into account the contribution of a positive cosmological constant, the following analyses are significant for the new model of universe.

		The Hubble parameter in terms of the scale function will be given by 

\begin{equation}
{H(a)}=\sqrt{{\beta \over a^3}+{\gamma \over a^4}+\lambda},
\end{equation}
while the cosmological energy density is described by the algebraic construction

\begin{equation}   \label{EDcosmo}
{\rho(a)}={3 \over \kappa}\left({\beta \over a^3}+{\gamma \over a^4}\right)
\end{equation}
with $\lambda={\Lambda \over 3}$ denoting the cosmological constant.                                      

The Friedmann equation for this model of universe has the form below:

\begin{equation}  \label{friedmancosmo}
{{\dot a }^2}={\beta \over a}+{\gamma \over a^2}+a^2 \lambda.
\end{equation}
To be noticed, as compared with the model treated in the previous section, the contribution brought by the cosmological constant to the Hubble function, while the cosmological energy density remains the same.
	
	After performing the variable separation, the differential equation (\ref{friedmancosmo}) becomes 

\begin{equation}  \label{diffeqfried}
{{ada} \over {\sqrt {\lambda \left(a^4+{\beta \over \lambda}a+{\gamma \over \lambda}\right)}} }=dt.
\end{equation}
At this point, by introducing the substitutions

\begin{equation}  \label{notparam}
p={\beta \over \lambda}, ~~~~~~~~~~~      q={\gamma \over \lambda},
\end{equation}
relation (\ref{diffeqfried}) transforms into 

\begin{equation}  \label{diffeqfried2}
{{ada} \over {\sqrt {\lambda}}\sqrt{a^4+pa+q}}=dt.
\end{equation}
If it is to consider that $a_i$ where $i=1..4$ are defining the four  roots of the quartic polynomial $a^4+pa+q$, then the differential equation (\ref{diffeqfried2})  can be rewritten as

\begin{equation}
{{ada} \over {\sqrt {\lambda}}\sqrt{(a-a_{1})(a-a_{2})(a-a_{3})(a-a_{4})}}=dt,
\end{equation}
leading after integration to the algebraic transcendental relation containing the roots $a_i$:

\begin{equation}  \label{transceneq}
{{{2} \over {\sqrt {\lambda}}\sqrt{(a_{1}-a_{4})(a_{2}-a_{3})}}\bigg[{a_{1}EllipticF[Z(a), \varepsilon]+(a_{2}-a_{1})EllipticPi[\varepsilon\prime, Z(a), \varepsilon]}}\bigg]=t,
\end{equation}
 with

\begin{equation} 
Z(a)=\arcsin{\left[ \sqrt{{(a-a_{2})(a_{1}-a_{4})} \over {({a-a_{1}})(a_{2}-a_{4})}} \right]},  {\varepsilon}= {{{(a_{1}-a_{3})(a_{2}-a_{4})}} \over {(a_{2}-a_{3})(a_{1}-a_{4})}}, \varepsilon\prime= {{a_{2}-a_{4}} \over {a_{1}-a_{4}}}.
\end{equation}
We want to stress that for $Z=0$, the two elliptic functions involved vanish. Our result can be put into correspondence with the one obtained by \cite{Aldrovandi:2005ya}, study in which a more generalized situation is analysed, namely the $\Lambda \gamma CDM$ Model (dust dark matter model).
Asymptotic representations for the elliptic integral of the first kind, $F[Z(a), \varepsilon]$ and the elliptic integral of the third kind, $Pi[\varepsilon\prime, Z(a), \varepsilon]$, allow for the transcendental equation (\ref{transceneq}) to reduce to a new, more simplified and algebraically convenient transcendental form:

\begin{equation}
{{{2Z(a)} \over {\sqrt {\lambda}}\sqrt{(a_{1}-a_{4})(a_{2}-a_{3})}}\bigg[{ a_{2}+{{a_{2}(\varepsilon+2\varepsilon\prime)-2\varepsilon\prime a_{1}} \over {6}}Z^2(a)}}\bigg]=t.
\end{equation}
An appropriate manner of tackling these transcendental equations are the numerical procedures. For more details into the behaviour of these elliptic functions, their asymptotic approximations, series expansions, inequalities, one can consult the papers \cite{Karp:2007sr} and references within.

We add the remark that the four roots $a_i$ are satisfying a set of mathematical relations in terms of the model parameters known as the Viete's relations.
	
		In what follows, we will discuss the nature of the four roots $a_i$  of the polynomial $a^4+pa+q$ and the expressions of the roots itself, this in the frame of the solutions of the quartic equation theoretical background. Thence, for the polynomial $a^4+pa+q$, its discriminant is:

\begin{equation}
\Delta=256q^3-27p^4,
\end{equation}
which, in terms of our notations in (\ref{notparam}), is equivalent with the algebraic expression

\begin{equation}   \label{deltacosmo}
\Delta={1 \over \lambda^3}\left(256\gamma^3-27{\beta^4 \over \lambda}\right).
\end{equation}
To be mentioned that the sign of the discriminant (\ref{deltacosmo}) has a significant influence over the nature of the roots $a_i$  with $i=1..4$. The analysis over this aspect can be refined through considering the sign of the polynomials below:

\begin{equation}  \label{polynomials}
P=0,    Q=8p=8{\beta \over \lambda}, D=64q=64{\gamma \over \lambda},  {\Delta_0}=12q={12\gamma \over \lambda}.
\end{equation}
Thus, according to the theory \cite{Rees:1922sr}, depending on the signs combination for the expressions defined in (\ref{polynomials}) together with the sign of the discriminant (\ref{deltacosmo}), one distinguishes a number of situations that determine the nature of the roots $a_i$ and their algebraic formulas. We will proceed to exposing the relevant situations for our polynomial under investigations.

\begin{itemize}

\item{ $\Delta<0$ }

	In this case, the quartic equation $a^4+pa+q=0$ possesses two real roots and two complex roots. The mathematical expression of the roots are:

\begin{equation}
a_{1,2}=-S\pm {1 \over 2}\sqrt{-4S^2+{p \over S}} \in \Re,   a_{3,4}=S+i\sqrt{S^2+{p\over{4S}}} \in \mathbb {C},
\end{equation}
where $S={1 \over 2\sqrt{3Q'}}\sqrt{Q'^2+12q}$, with $Q'=\root 3\of{{\Delta_1+{\sqrt{{\Delta_1}^2-4{\Delta_0}^3}} \over2}}$ and $\Delta_1=27p^2=27{\beta^2 \over \lambda^2}$, $\Delta_0=12q=12 {\gamma \over \lambda}$.
Because the scale factor is a real quantity, particularly, we will be interested in the real roots.
	
By virtue of identity ${\Delta_1}^2-4{\Delta_0}^3=-27\Delta= 27^2{\beta^4 \over \gamma^4}-6912{\gamma^3 \over \lambda^3}$, we have that

\begin{equation}
{Q'}=\root 3\of{{\Delta_1+{\sqrt{{\Delta_1}^2-4{\Delta_0}^3}} \over2}}=\root 3\of {\Delta_1+{\sqrt{-27\Delta}} \over 2}>0,\nonumber\\
\end{equation}
where the discriminant $\Delta$ in terms of the model parameters is given by (\ref{deltacosmo}).
In these considerations, the real solutions are described by the algebraic expression

\begin{equation}
a_{1,2}=-{1 \over {2\sqrt{3Q'\lambda}}} \sqrt{\lambda Q'^2+12\lambda} \pm {1 \over 2}\sqrt{-{1 \over{3Q'\lambda}}(\lambda Q'^2+12\gamma)+{{2\beta\sqrt{3Q'}} \over {\sqrt{\lambda}\sqrt{\lambda Q'^2+12\gamma}}}},
\end{equation}
where the formula for $Q'$  has been presented above.

	In addition to this, by calling down the inequality $\Delta <0$, we find that the parameters of our model are subjected to the constraint:

\begin{equation}
\gamma^2 \lambda<0,1 {\beta}^4.
\end{equation}

\item{ $\Delta>0$ }
	
	As for our model  $P=0$, we find out that this situation is not reflected in theory. 

\item{ $\Delta=0$ }

	In this case, by considering the condition $D=0$, there are identified as solutions one real double root and two 	complex ones.

	In the same time we have  ${\Delta_1}^2-4{\Delta_0}^3=0$ which leads to the parametric relation between the radiative and cosmic dust components:

\begin{equation}
{\beta^4 \over {\gamma^3 \lambda}}={256 \over 27}\simeq 9.4. 
\end{equation}

The real solutions will be described by the same algebraic construction in (39), with $Q'={{3{p^{2/3}} \over {2^{1/3}}}={3 \over 2^{1/3}}\left({\beta \over \lambda}\right)^{2/3} }$ which as a result of a trivial calculus leads to the following set of real solutions:

\begin{equation}
a_1=a_2=-\left({q \over 3}\right)^{1/4}=-\left(\gamma \over {3\lambda}\right)^{1/4}. 
\end{equation}
One can also tackle equation (\ref{transceneq})  by making use of numerical procedures and analyses which might provide other new insights into the Einstein solutions of the model.

\end{itemize}
	Before we finish this section of our study, from the first Einstein equation (\ref{einstein2}) we derive the evolution of pressure with respect to the scale function:

\begin{equation}
p(a)={\gamma \over {\kappa{a^4}}}-{{3\lambda} \over \kappa}. 
\end{equation}
This result shows that the effect of the cosmological constant is to reduce the pressure. Furthermore, with recalling the energy density (\ref{EDcosmo}), we are able to determine a form of effective pressure dependence with respect to the energy density where a scale factor ‘interference’ is to be noticed:

\begin{equation}  \label{polytropic}
p(\rho, a)=\rho-{1 \over \kappa}\left({3\beta \over a^3}+{2\gamma \over a^4}+3\lambda\right). 
\end{equation}
We note that the non-linear, polytropic equation (\ref{polytropic}) is the algebraic sum of a standard linear EoS $(p\simeq \alpha\rho c^2)$ and a non-linear term dependent on all the parameters’ model. With respect to the linear dependence, we point out that within this level of approximation (the ultra-relativistic limit), it might be viewed as a ‘version’ (we used commas since we do not have a ‘pure’ $p(\rho)$ EoS) of a stiff equation of state ($p \simeq \rho$), the last one brought into evidence by the Zel’dovich pioneering model of Universe which assumes that in the early stage of Universe, in the vicinity of the cosmological singularity, the universe composition is described by a cold gas of baryons interacting through a meson field \cite{ Zel’dovich:1972sr}. A similar non-linear EoS has been obtained by us for a mixed cosmology with stiff fluid, cosmic dust and a cosmological constant \cite{Dariescu:2016sr} in the context of fluid dynamics with viscous effects. More precisely, for that model, we found out that a specific relation between the model parameters could make the non-linear contribution factorizable so that a bulk viscosity coefficient has been identified. 

\section{Conclusions}

	Within the frame of a quadri-dimensional FRW metric with zero curvature, we analysed the dynamics of a cosmology where the universe is filled with a mixture of matter species namely cosmic dust and radiation. A similar analytical endeavour has been considered by Chavanis \cite{CHAVANIS:2015sr} who offers a succint analysis that presents the cubic equation governing the evolution in time of the scale factor together with some reports on the behaviour of the scale factor and total energy density both in the primordial state ($a(t=0)=0$) and far away in the future ($t=+\infty$). In this respect, our result has matched the one obtained by Chavanis  \cite{CHAVANIS:2015sr}. Beside these, our investigations have gone far beyond. We offered a discussion on the nature of the solutions of the cubic equation $-$ whose variable is the scale factor that can admit an interpretation in terms of an exact solution of Einstein's field equations (\ref{einstein2}) $-$ by considering the mathematical theoretical background of the cubic equation.
Morever, it was discovered that a radiative epoch characterizes the early times of the cosmogenesis in agreement with all the cosmologies developed within the null- primordial conditions $a(t=0)=0$. On the other hand, very far away in the future, it was identified that a dust-like behaviour dominates the universe dynamics. In addition to this, we have included the thermodynamics of the model and also the inverse function of the EoS. As a feature of this universe we found that it is decelerating as the positivity of the acceleration parameter (\ref{accparam_a}) indicates. The inverse function of the EoS, the universe constant entropy and the temperature dependences of the pressure, cosmological energy density and scale factor were determined. As for the entropy. it was found that it depends on the lower threshold of temperature and on the parameter responsible for the dust component. Also, it proves that as the scale factor approaches the primordial state ($a\to0$), the energy density diverges $\rho \to \infty$. Very far away in the future, the Hubble function possesses an asymptotic behaviour as $H \sim 2/{3t}$, the scale factor admits an Einstein de-Sitter time dependence, the temperature of the universe decreases until it reaches its minimum value $T_{0}$, while the total energy density and the pressure approach the zero value. 
	
	When a cosmological constant is manifesting its presence within the model, one prime effect is to reduce the cosmic pressure. A non-trivial dependence of the pressure with respect to the energy density and the scale factor has been deduced in terms of an algebraic sum between a Zel'dovich stiff matter linear EoS component and a non-linear term involving the model' parameters.

\end{document}